\documentclass[review]{elsarticle}

\usepackage{color}

\usepackage{amsmath}
\usepackage{hyperref}

\journal{arXiv.org}

\begin{document}

\begin{frontmatter}

\title{Fine-tuning the etch depth profile via dynamic shielding of ion beam}

\author{Lixiang Wu}
\ead{wulx@mail.ustc.edu.cn}

\author{Keqiang Qiu}
\ead{blueleaf@ustc.edu.cn}

\author{Shaojun Fu}
\ead{sjfu@ustc.edu.cn}

\address{National Synchrotron Radiation Laboratory, University of Science and Technology of China, Hefei 230029, China}

\begin{abstract}
We introduce a method for finely adjusting the etch depth profile by dynamic shielding in the course of ion beam etching (IBE), which is crucial for the ultra-precision fabrication of large optics.
We study the physical process of dynamic shielding and propose a parametric modeling method to quantitatively analyze the shielding effect on etch depths, or rather the shielding rate, where a piecewise Gaussian model is adopted to fit the shielding rate profile.
We have conducted two experiments.
In the experiment on parametric modeling of shielding rate profiles, its result shows that the shielding rate profile is significantly influenced by the rotary angle of the leaf.
And the experimental result of fine-tuning the etch depth profile shows good agreement with the simulated result, which preliminarily verifies the feasibility of our method.
\end{abstract}


\end{frontmatter}


\section{Introduction}
\label{sec-intro}

As a state-of-art technique, IBE plays a crucial role in the ultra-precision fabrication of large optics such as X-ray mirrors \cite{peverini2010ion}, vacuum ultraviolet gratings \cite{xu_vuv_2005}, and beam sampling gratings (BSGs) \cite{rao2014chemical,wu_algorithms_2015}.
Collimators or shutters are usually used to adjust the etch rate profile \cite{peverini2010ion,savvides2006correction} in order to meet the desired etch depth map.
L. Peverini \textit{et al} presented an IBE technique using a double-blade system to profile flat optics into aspherical X-ray mirrors \cite{peverini2010ion}.
In the fabrication of BSGs, we also encounter a similar scenario and need to adjust the spatial distribution of etch depths.
But in contrast to the surfacing of X-ray mirrors, the adjustment of etch depths in the fabrication of BSGs is, to some extend, more difficult and requires more flexibility \cite{rao2014chemical,wu_algorithms_2015,wu_finely_2015}.
On one hand, the etch depths are almost randomly distributed.
On the another, the adjustment of etch depths is quite fine and slight compared with the maximum etch depth.
Hence, in our previous work \cite{wu_algorithms_2015,wu_finely_2015}, we proposed a new solution for fabricating the large-aperture BSG of high diffraction uniformity as well as algorithms for finely tuning the spatial distribution of etch depths.

To put the idea into practice, we investigate the physical process of dynamic shielding and introduce how to calculate the shielding time.
With a graphite leaf, the ion beam emitted from a linear ion source is locally blocked.
The leaf scans along the major axis of the ion source and simultaneously rotate about itself, then the ion beam impinging on the substrate to be etched is dynamically adjusted.
We also conduct quantitative analysis on the shielding effect on etch depths, or rather the shielding rate.
A piecewise Gaussian model is adopted to fit the shielding rate profile.
Finally, experiments are conducted for parametric modeling of the shielding rate profile and verifying the feasibility of fine-tuning the etch-depth profile.

\section{Dynamic shielding during ion beam etching}
\label{sec-methods}

\subsection{Shielding rate profile}

\begin{figure}[htbp]
\centering
\includegraphics[width=\columnwidth]{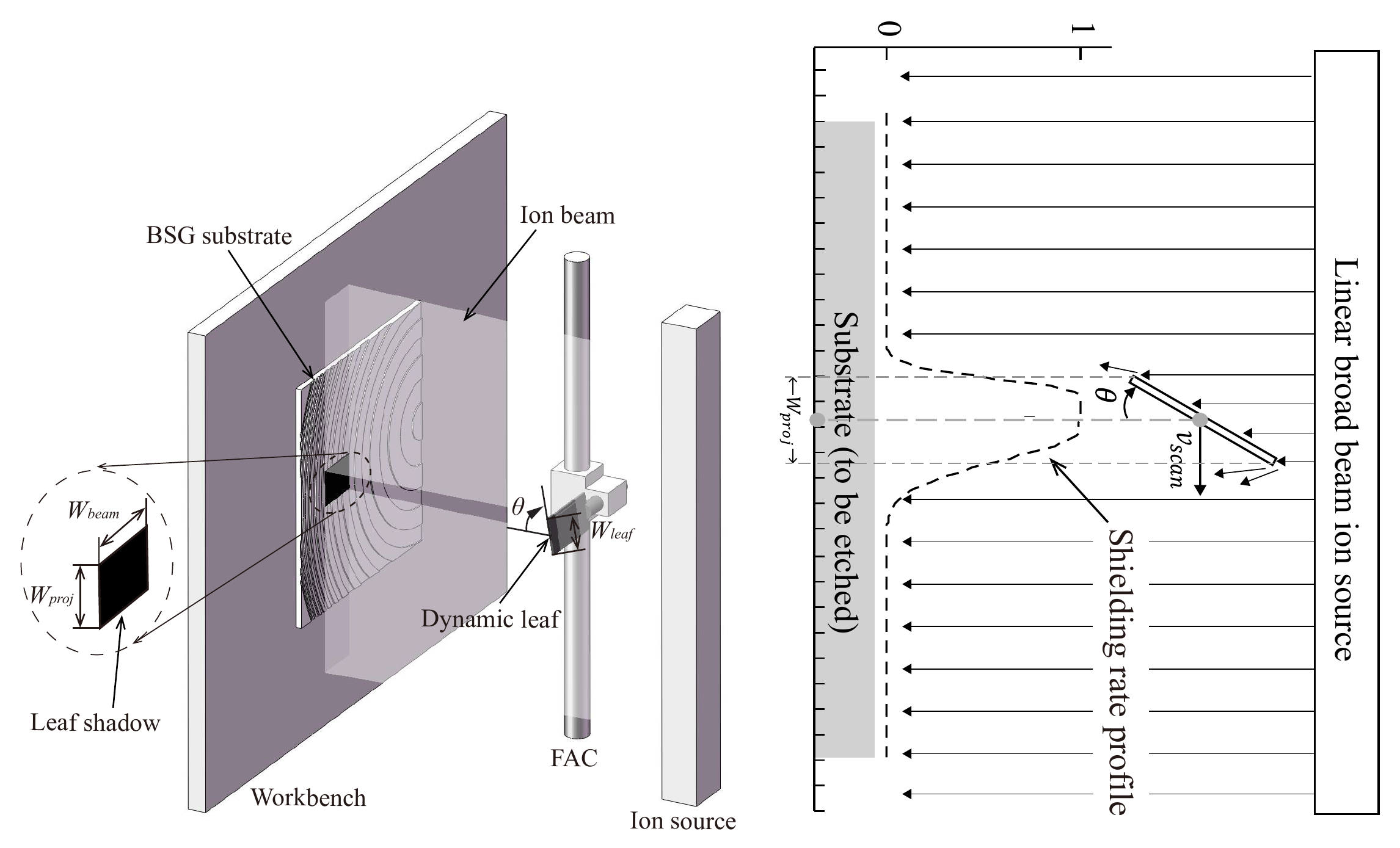}
\caption{Dynamic shielding during IBE. The left illustrates the IBE of BSG and the right is the abstract model in 2D. Where $W_{proj}$ is the projected width, $W_{beam}$ is the beam width, $W_leaf$ is the leaf width, $\theta$ is the rotary angle of the dynamic leaf, and $v_{scan}$ is the constant scanning velocity.}
\label{fig:IBEwithDynamicLeaf}
\end{figure}

A dynamic leaf, which is a rectangular thin graphite plate, has been designed and is dragged by stepper motors. As illustrated in Figure~\ref{fig:IBEwithDynamicLeaf}, the dynamic leaf has two degrees of freedom, i.e., scanning vertically and swinging about itself within a limited angular range.
During the etch process, the dynamic leaf locally blocks the ion beam and then a shadow is formed on the substrate.
To quantitatively describe the shadow, the shielding rate profile is introduced (see Fig.~\ref{fig:IBEwithDynamicLeaf}), where the shielding rate reflects the shielding effect on protecting the materials from etching.
The shielding rate of 0 indicates no shielding and that of 1 indicates full shielding and being free of etching.

The full-width-half-maximum (FWHM) of the shielding rate profile is defined as the shadow width, which is different from the projected width of the dynamic leaf.
As is shown in Figure~\ref{fig:IBEwithDynamicLeaf}, the projected width can be calculated by
\begin{equation}
W_{proj} = W_{leaf}\sin{\theta} + H_{leaf}\cos{\theta},
\end{equation}
where $W_{proj}$ is the projected width, $W_{leaf}$ is the leaf width, $H_{leaf}$ is the leaf thickness, and $\theta$ is the rotary angle of the dynamic leaf.
Note that, in this paper, we focus on the 1D analysis of the leaf shadow and the shielding rate profile, so it is assumed that the beam width $W_{beam}$, as illustrated in Figure~\ref{fig:IBEwithDynamicLeaf}, has nothing to do with the shielding rate.


A piecewise Gaussian model is applied to fitting the shielding rate profile, which is given by 
\begin{equation}
PWG(x;a,b,c_1,c_2)=\begin{cases}
  ae^{-(\frac{x+\frac{b}{2}}{c_1})^2} & \text{for $x < -\frac{b}{2}$} \\
  ae^{-(\frac{x-\frac{b}{2}}{c_2})^2} & \text{for $x > \frac{b}{2}$} \\
  a & \text{otherwise}
\end{cases},
\end{equation}
The FWHM of the piecewise Gaussian is defined as
\begin{equation}
FWHM\{PWG(x;a,b,c_1,c_2)\} = b + \sqrt{\ln{2}}(c_1 + c_2),
\end{equation}
where $\sqrt{\ln{2}}c_1$ and $\sqrt{\ln{2}}c_2$ are the FWHMs of the left half-Gaussian and the right half-Gaussian.

\subsection{Calculation of the shielding time}

\begin{figure}[htbp]
\centering
\includegraphics[width=.8\columnwidth]{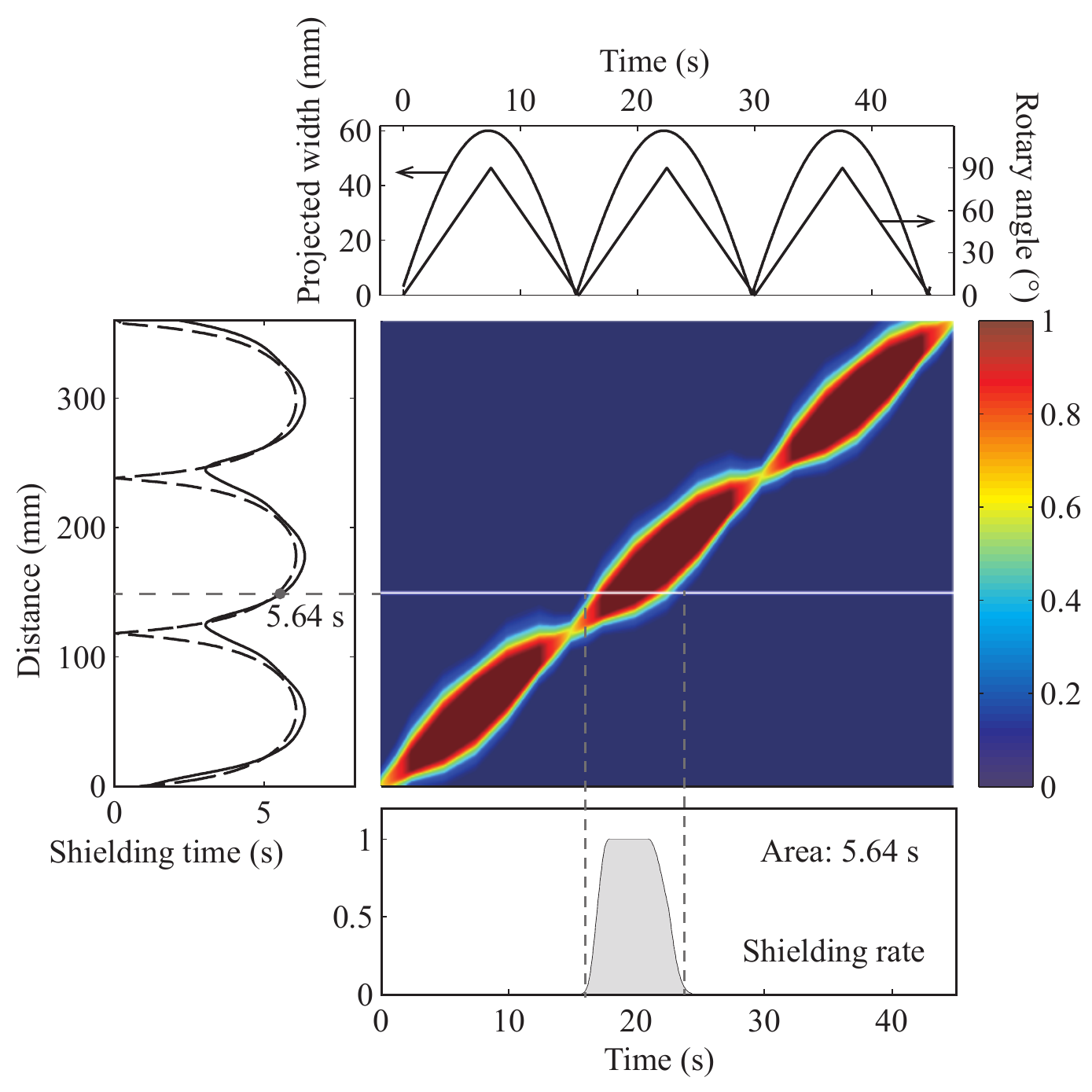}
\caption{Calculation of shielding time. The plots at the top show the rotary angle profile and the corresponding projected width profile. The lower plot shows the shielding rate profile at the distance of $150~mm$ and the calculated shielding time is $5.64~s$. The left part presents two shielding time profiles, of which the solid curve represents the actual result and the dashed curve represents the approximate result.}
\label{fig:ShieldingTimeCal}
\end{figure}

The shielding time of a point herein refers to the total elapsed time the point is shielded by the dynamic leaf.
For clarity, the term, dwell time used in the previous paper \cite{wu_algorithms_2015}, is renamed as shielding time hereafter.
To calculate the shielding time, we should map the shielding rate profiles onto the spatial temporal coordinate system and then, as shown in Figure~\ref{fig:ShieldingTimeCal}, integrate the fitted shielding rate profile over time.
The plots at the top of Figure~\ref{fig:ShieldingTimeCal} show the rotary angle profile and projected width profile, which, accompany with the constant scanning velocity, determines the motion trajectory of the dynamic leaf, i.e., scanning at a constant speed and swinging between $0^\circ$ and $90^\circ$ simultaneously.
The left part of Figure~\ref{fig:ShieldingTimeCal} presents the calculated shielding time profiles, of which the solid curve is the actual shielding time profile and the symmetric dashed curve indicates the approximate result assuming that full shielding occurs in the project area and other areas are free of shielding.
Obviously, the actual shielding time profile is asymmetric due to the asymmetry of the shielding rate profile, which will be discussed in the following section.

\section{Results and discussions}
\label{sec-results}

\subsection{Parametric modeling of shielding rate profiles}
\label{subsec-edge-scattering}

\begin{figure}[htbp]
\centering
\includegraphics[width=\columnwidth]{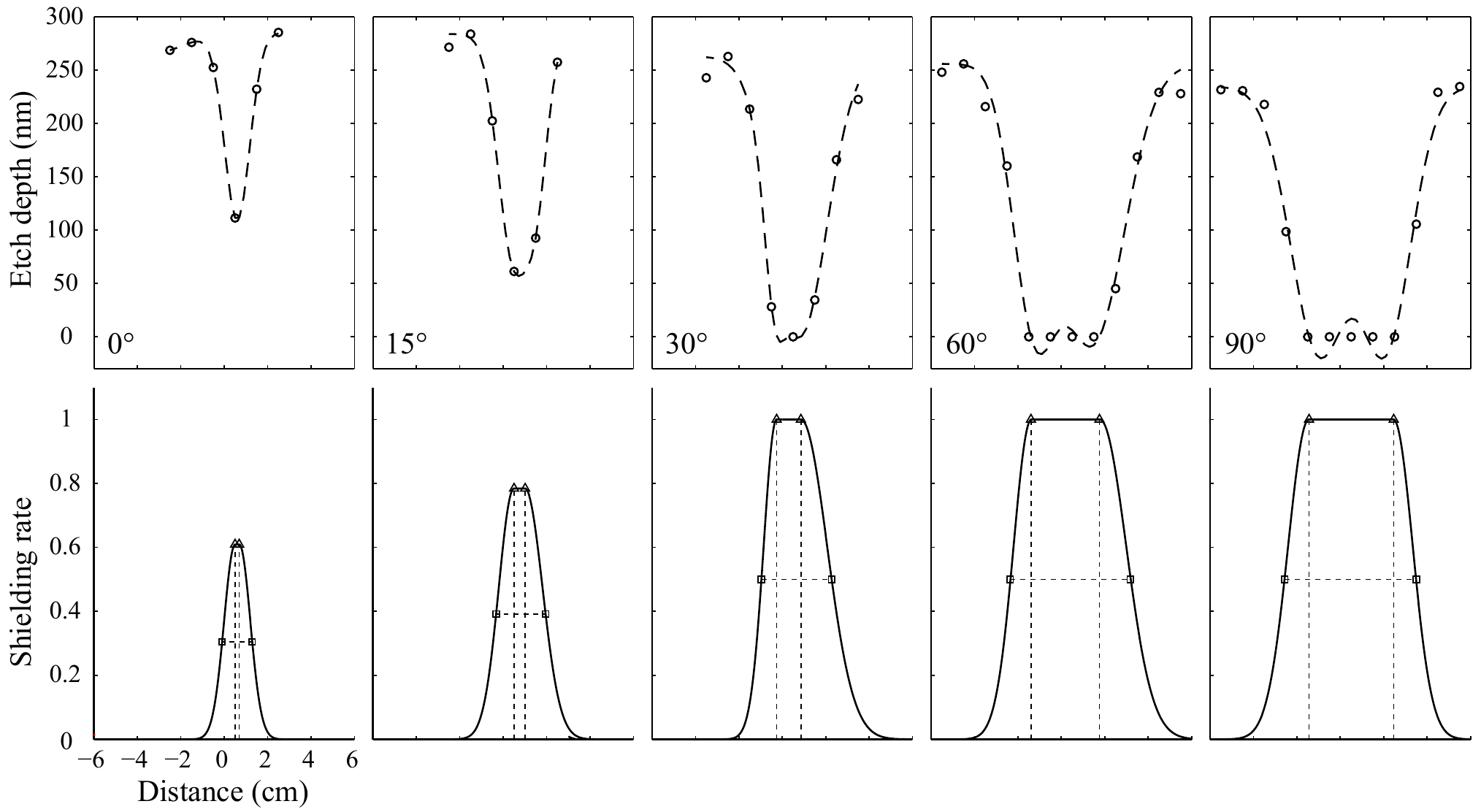}
\caption{The measured etch depth profiles and the fitted shielding rate profiles at different rotary angles. The upper five plots show the measured etch depth profiles with the rotary angles of $0^\circ$, $15^\circ$, $30^\circ$, $60^\circ$, and $90^\circ$, respectively and the corresponding plots in the lower half illustrate the shielding rate profiles, which fit the piecewise Gaussians (the solid curves). Note that the circles in the upper five plots represent the experimental results.}
\label{fig:EtchDepthShieldingRate}
\end{figure}

We have observed that the shielding rate profile is influenced by the rotary angle of the dynamic leaf. 
To further quantify its influence on the distribution of etch depths, we conducted a well-designed experiment, which consists of 5 groups in correspondence with rotary angles of $0^\circ$, $15^\circ$, $30^\circ$, $60^\circ$, and $90^\circ$, respectively.
A number of pieces of fused quartz samples are prepared for the etching experiment and the etch depth is measured by a stylus profiler (AMBIOS XP-1) with vertical resolution of the order of 0.1 nm.
In Figure~\ref{fig:EtchDepthShieldingRate}, the upper half shows the measured etch depth profiles at different rotary angles and the lower half illustrates the corresponding shielding rate profiles, which are asymmetric when the rotary angle is not equal to $0^\circ$ or $90^\circ$.
The shielding rate of ion beam, which indicates the blocking effect of ion beam for the dynamic leaf, is inversely proportional to the measured etch depth.

\begin{figure}[htbp]
\centering
\includegraphics[width=.7\columnwidth]{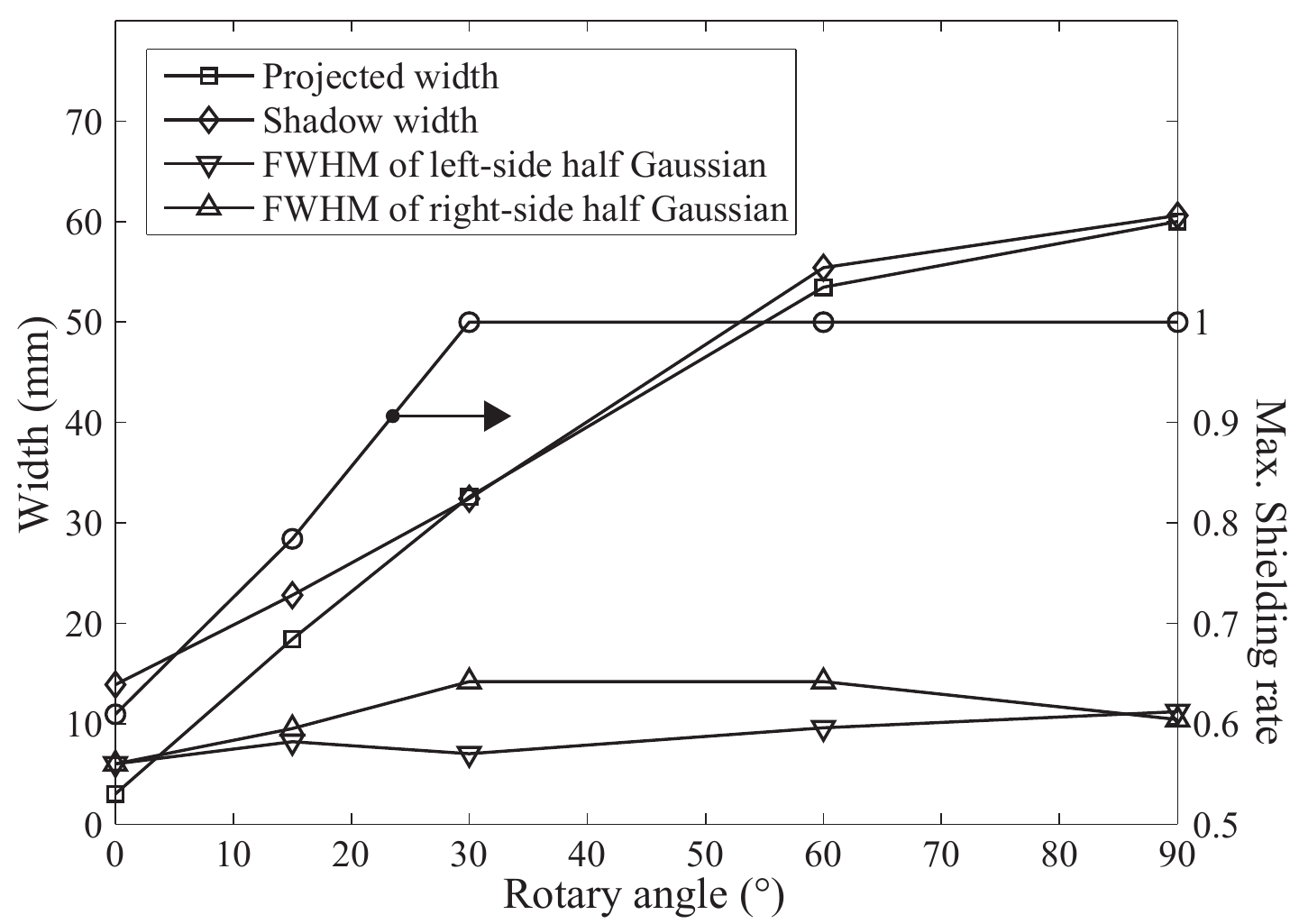}
\caption{Summary of the parametric model. The left Y axis shows the projected width and measured FWHM with different rotary angles and the right part shows the shielding rate with different rotary angles.}
\label{fig:ParametricModel}
\end{figure}

The Monte Carlo simulation \cite{gorelick_aperture-edge_2009,gorelick_aperture-edge_2009-1}, as is well known, is a general approach to analyze ion scattering from the leaf edge and its effect on etch depths.
Alternatively, parametric modeling we adopted here is a more practical way to study the leaf-edge scattering, especially for studying its effects on etch depths.
Based on the above fitted results and the piecewise Gaussian model, 
a parametric model of shielding rate profiles is built, which is determined by 4 parameters including the shadow width, the FWHM of the left half-Gaussian, the FWHM of the right half-Gaussian, and the maximum shielding rate.
Figure~\ref{fig:ParametricModel} shows that the parameters vary with the rotary angle: the shadow width and the projected width increase with the rotary angle but the shadow width is larger than the projected width when the rotary angle is within $30^\circ$; the FWHM of right half Gaussian is larger than that of left half Gaussian; the maximum shielding rate with rotary angle that is larger than around $30^\circ$ keep stable on $1$.
Note that the leaf width is specified as 60~mm. 

We found that the proximal side of the dynamic leaf that is closer to the substrate gives rise to the narrow FWHM, e.g. the FWHM of the left half Gaussian in Fig.~\ref{fig:ParametricModel}.
In summary, the rotary angle of the dynamic leaf accounts for the asymmetry of the two half Gaussians and, with this parametric model, the shielding rate profile at arbitrary rotary angle within $[0^\circ, 90^\circ]$ can be calculated by linear interpolation.

\subsection{Fine-tuning the etch depth profile}

\begin{figure}[htbp]
\centering
\includegraphics[width=.55\columnwidth]{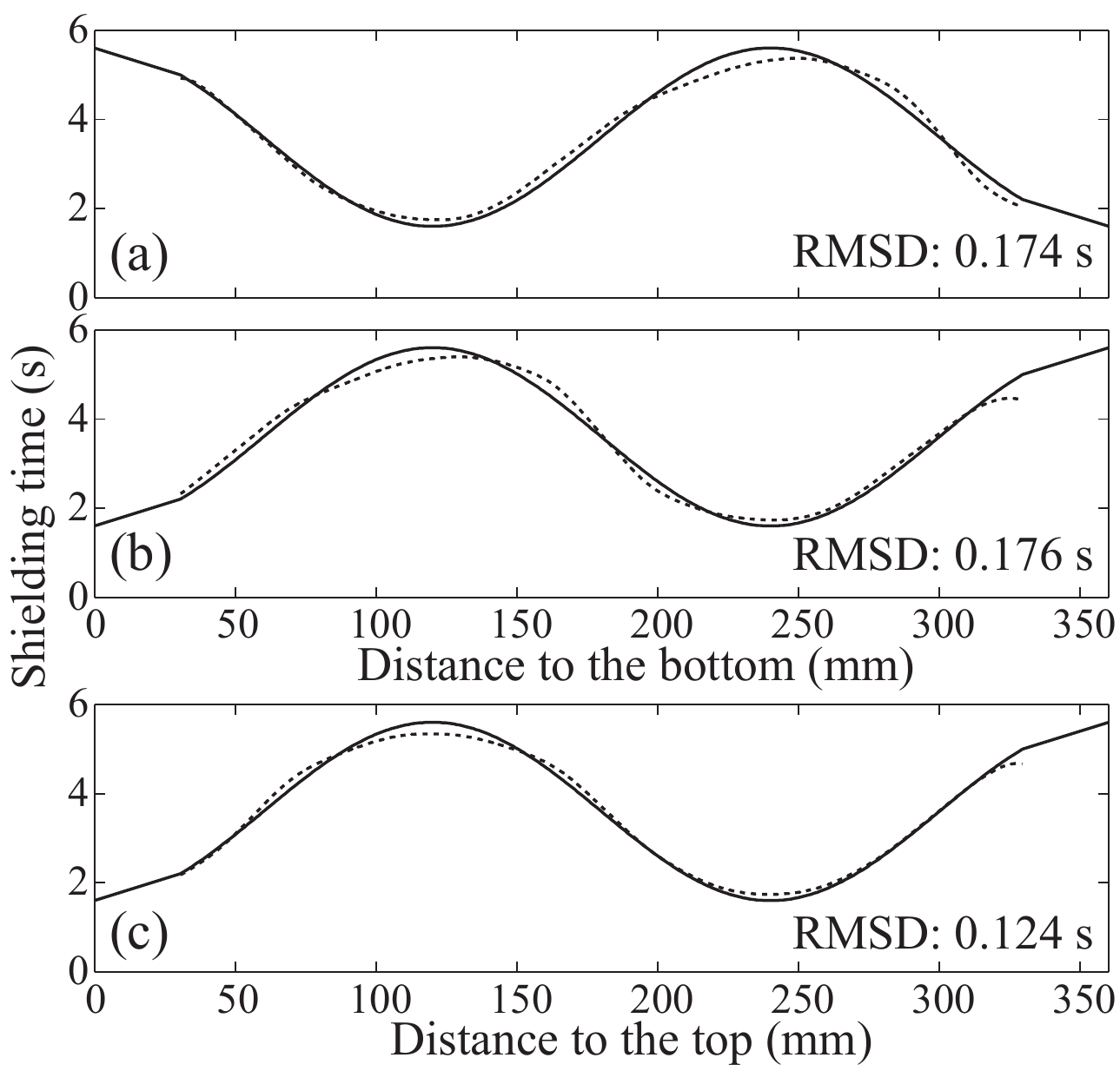}
\caption{The optimization of shielding times. (a) the shielding time optimization for a bottom-up stroke; (b) the shielding time optimization for a top-down profile, which is actually a bottom-up stroke when the substrate is vertical flipped; (c) the average of (a) and (b). Note that the solid line indicates the target shielding time profile and the dashed line represents the optimized shielding time profile. it repeats 20 rounds of strokes in total.}
\label{fig:OptShieldingTimes}
\end{figure}

We have designed an experiment to verify the feasibility of fine-tuning the etch depth profile.
Typically, a sinusoidal shielding time profile is specified as the target and then the motion trajectory of the dynamic leaf is optimized.
A number of fused silica substrates are prepared for evaluating the etch depth profile, which is measured by the aforementioned high resolution stylus profiler.
However, it is difficult to optimize the shielding time profile that fits the symmetric sinusoidal profile because of the asymmetry of the shielding rate profile.
For that, we optimized two shielding time profiles, i.e., the bottom-up profile and the top-down profile, and the summation of the two profiles shows better agreement with the target sinusoidal profile (see Figure~\ref{fig:OptShieldingTimes}).

\begin{figure}[htbp]
\centering
\includegraphics[width=.6\columnwidth]{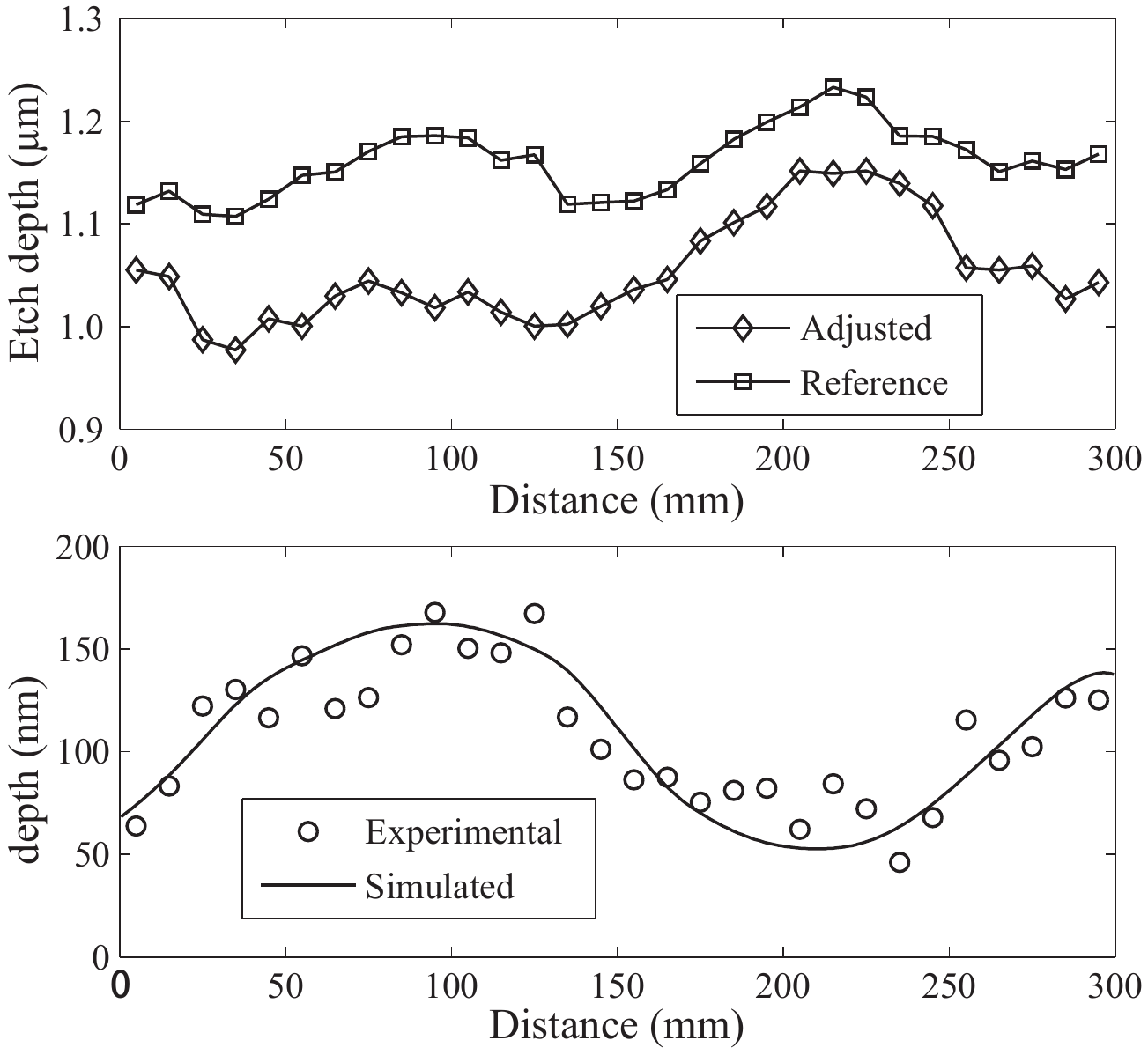}
\caption{Comparison between the experimental result and the simulated prediction. The upper half shows the reference etch depths and the etch depths after fine adjustment; the lower half shows the comparison.}
\label{fig:ComparisonExpSim}
\end{figure}

In this experiment, a reference group is set to avoid the etch depth variations caused by the spatial non-uniformity of the etch rate.
The depth difference between the adjusted profile and the reference profile is the fine-tuned etch depth profile.
Figure~\ref{fig:ComparisonExpSim} presents that the experimental result shows agreement with the simulated or optimized result in trend.
However, there are still significant deviations in several points.
Further investigation on depth deviation would be conducted in the future work.

\section{Conclusion}
\label{sec-conclusion}

We have introduced a method for fine-tuning etch depths during ion beam etching and have conducted experiments to verify the feasibility of the method.
In the first experiment, we investigated the relationship between the rotary angle and the etch depth profile by parametric modeling.
In the second, the comparison between the simulated result and the measured result shows that fine-tuning the etch depth profile is feasible.
It is worth mentioning that the method for fine-tuning etch depths via dynamic shielding of ion beam not only can be applied in the fabrication of BSGs but also would be promisingly used in the ultra-precision fabrication of other large optics.
The future work will focus on the experimental realization of fine-tuning etch depths on the 2D plane.

\section*{Acknowledgement}
\noindent We would like to thank Yves Daoust for answering my question on math.stackexchange.com. The work was partially supported by the National Natural Science Foundation of China under Grant No. 11275201.

\section*{References}

\bibliography{references}

\end{document}